\documentclass{article}
\usepackage[centertags]{amsmath}
\usepackage{amsfonts,latexsym}
\usepackage{amssymb}
\usepackage{amsthm}

\newcommand{\be}{\begin{equation}}
\newcommand{\ee}{\end{equation}}
\newcommand{\bea}{\begin{eqnarray}}
\newcommand{\eea}{\end{eqnarray}}

\def\pint{\mathbin{\raise1.5pt\hbox{$\underline{\raise-.5pt\hbox
{$\phantom{n}$}}\mskip-2mu\scriptstyle|$}}} \large
\def\th{\mathop{\rm th}\nolimits}
\def\tg{\mathop{\rm tg}\nolimits}
\def\ctg{\mathop{\rm ctg}\nolimits}

\begin{document}
\begin{flushright}
hep-th/0401022
\end{flushright}

\begin{center}
\Large{\bf Wick quantization  of cotangent bundles over
Riemannian manifolds} \\
\vspace{1.0cm}

\large{I.V. Gorbunov\footnote{E-mail: ivan@phys.tsu.ru},
S.L. Lyakhovich\footnote{E-mail: sll@phys.tsu.ru} and
A.A. Sharapov\footnote{E-mail: sharapov@phys.tsu.ru}
}\\[2mm]

\footnotesize{{\it Physics Faculty, Tomsk State University,
Lenin Ave. 36, Tomsk 634050, Russia} } \\
\end{center}

\begin{abstract}
A simple geometric procedure is proposed for constructing Wick
symbols on cotangent bundles of Riemannian manifolds. The main
ingredient of the construction is a method of endowing the
cotangent bundle with a formal K\"ahler structure. The formality
means that the metric is lifted from the  Riemannian
ma\-ni\-fold~$\mathcal Q$ to its phase space $T^\ast{\mathcal Q}$ in the
form of formal power series in momenta with the coefficients being tensor
fields on the base. The corresponding K\"ahler two-form on the total
space of $T^\ast{\mathcal Q}$ coincides with the canonical symplectic
form, while the canonical projection of the K\"ahler metric on the base
manifold reproduces the original metric. Some examples are considered,
including constant curvature space and nonlinear sigma models,
illustrating the general construction.
\end{abstract}
{\it PACS:} 02.40.Tt; 
04.60.Ds; 
03.70.+k; 
03.65.Ca\\  
{Keywords:} Deformation quantization; K\"ahler geometry; Sigma-models.

\section{Introduction}

The word ``quantization''  usually means  a procedure of
constructing a quantum mechanical system for a given classical
one. This construction is known to involve a great amount of
ambiguity and, in this sense, the quantum mechanics provides a
more refined description of physical systems than the classical
mechanics. Mathematically, this deeper level of description
manifests itself in extra geometric structures to be defined on
the phase space of a system to perform its consistent
quantization. In the framework of the deformation quantization,
for instance, a symplectic connection, being not involved in the
classical dynamics, becomes a key ingredient of the theory at the
quantum level \cite{Fedosov1}, \cite{Fedosov2}. Special types of
the deformation quantization can involve more ``rigid''
geometrical structures (like metric, torsion, complex structure,
etc.) and different choices for these structures may lead to
inequivalent quantizations of the same phase space.

In this paper we address the question of the Wick deformation
quantization on cotangent bundles of Riemannian manifolds.
Several reasons can be mentioned motivating this study:

{(i)} The phase space of most physical systems has the form of
cotangent bundle $T^\ast{\mathcal Q}$ over the configuration space
$\mathcal Q$. Often, the latter carries a natural
(pseudo-)Riemannian metric entering the very formulation of
the classical model (point particle in the General Relativity,
nonlinear sigma-models, etc.). Even when such a metric is not
explicitly involved at the classical level it may appear
(sometimes implicitly) upon quantization. For example, the
configuration space metric is a basic ingredient of the unique
effective action construction \cite{vilk}, \cite{barv-vilk}, the
most advanced path-integral quantization method known in the field
theory. In the series of papers \cite{K1}, \cite{K2}, \cite{KSh},
the relevance of phase-space metric was argued to the
regularization of phase-space path integrals, including those
modelling transition amplitudes in quantum gravity \cite{WK}.

{(ii)} It is the Wick symbol algebra of physical observables (the
Bargmann-Fock representation) which is commonly accepted to serve
as a basis for the quantum mechanical description of fields.
Usually, the Wick symbols are known but at the level of free
fields, that suggests the perturbative treatment for the
interaction from the outset. Such a disintegration of the entire
theory into the (linear) ``free part'' and the (nonlinear)
``interaction'' may happen to be inadequate to the physics, as it
can break,  for example, fundamental symmetries of the classical
model. The nonlinear sigma-models and, in particular, strings on
the AdS space are typical examples where no physically reasonable
linear approximation can be found for the phase/configuration
space geometry. Thus, to get a quantum mechanical description
respecting the geometry of such essentially nonlinear models, like
just mentioned, one needs an explicitly covariant and globally
defined procedure of Wick quantization for cotangent bundles.

{({iii})} Finally, the construction of Wick symbols is intimately
connected with the metrization of a phase space to be quantized.
In  fact, whenever all the conditions on the metric, imposed by
the Wick quantization, are satisfied, the phase space is proved to
be the  K\"ahler manifold~\cite{DLSh}. The bundle structure of the
phase space  $T^\ast{\mathcal Q}$ allows one to endow the
configuration space $\mathcal Q$ with the Riemannian metric
induced  by the canonical embedding ${\mathcal Q}\subset
T^\ast{\mathcal Q}$.

The main idea of this paper  is, in a sense, an inversion of the
last remark. Starting with the Riemannian metric $g$ on the
configuration space $\mathcal Q$, we construct a {\it formal} K\"ahler
metric
\begin{displaymath}\displaystyle
G=G_{\imath\bar\jmath}dz^\imath d\bar z^\jmath =
G_{ij}(x,p)dx^idx^j + G_{i}^{j}(x,p)dx^idp_j + G^{ij}(x,p)dp_idp_j
\end{displaymath}
on the phase space $T^\ast{\mathcal Q}$ requiring the corresponding K\"ahler
two-form to coincide with the canonical symplectic structure on
$T^\ast{\mathcal Q}$
\begin{displaymath} \displaystyle
\omega=G_{\imath\bar\jmath}dz^\imath\wedge d\bar
z^\jmath=dp_i\wedge dx^i \, .
\end{displaymath}
Here $(z^\imath, \bar z^\jmath)$ are the complex coordinates
adapted to the K\"ahler structure, while $(x^i, p_j)$ are the
canonical coordinates on the cotangent bundle. In this context,
the formality means that the components of the metric tensor $G$
are given by formal power series in the momenta $p_i$.  When
$({\mathcal Q},g)$ is a real-analytical Riemannian manifold, these
series can be shown to converge in a tubular neighborhood of
${\mathcal Q}\subset T^\ast{\mathcal Q}$.  In this paper we
construct a natural phase-space metric $G$ satisfying boundary
condition $G|_{\mathcal Q}=g$ and equipping $T^\ast{\mathcal Q}$
with an integrable K\"ahler structure. Such a lift of the metric
from the base manifold to the cotangent bundle automatically
endows the latter with a (formal) connection respecting both the
canonical Poisson bracket and the phase-space metric.

Once the K\"ahler structure is defined, known deformation
quantization methods can be immediately applied to build up the
algebra of Wick symbols \cite{BW}, \cite{DLSh}, \cite{KS}.

The Weyl deformation  quantization of cotangent bundles has been
already studied in the Refs.\ \cite{BNW1}, \cite{BNW2}. In those
papers an affine connection has been lifted from the configuration
space $\mathcal Q$ to the so-called homogeneous connection on the
phase space $T^\ast{\mathcal Q}$ which is used then to apply the
machinery of the Fedosov deformation quantization \cite{Fedosov1,
Fedosov2}. In fact, the requirement of homogeneity restricts the
components of the connection to be at most linear in momenta. The
Wick quantization, as we will see in this paper, implies a
different lift for the connection (as it has to respect the
metric), leading in general to infinite series in momenta for the
lifted connection.

The paper is organized as follows. For reader's convenience, in
Sec.~2 we collect some basic notions and formulae related to the
K\"ahler geometry. In Sec.~3 a general covariant procedure is
proposed for the lift of a Riemannian metric on $\mathcal Q$ to a
metric on $T^\ast{\mathcal Q}$ equipping the latter with a
K\"ahler structure. This procedure is quite similar to the
construction of a formal exponential map of Ref.~\cite{EW}
inspired, in turn, by the ``flattering'' procedure for the
quantum connection in the Fedosov deformation quantization. In
all the cases one deals with an iterative construction of formal
power series in the fiber coordinates. In Sec.~4 a set of
holomorphic coordinates compatible with the K\"ahler structure
is obtained via the exponential map generated by a Hamiltonian
flow. The construction makes it possible to prove the
convergence of the formal series for the lifted metric in a
tubular neighborhood of the base $\mathcal{Q}\subset
T^\ast{\mathcal Q}$, provided $({\mathcal Q},g)$ is
real-analytical. In Sec.~5 we calculate the first Chern class of
the K\"ahler structure which turns out to be zero. In Sec.~6 the
general method is applied to several examples of physical
interest including those related to the quantum field theory. In
Section 7 we briefly discuss the issues of equivalence between
Wick and Weyl quantization, and problems of applying the
proposed technique to the field/string theory models.

\section{K\"ahler manifolds}
\setcounter{equation}{0}

In this section we briefly recall some basic definitions and facts
concerning the geometry of (almost-)K\"ahler manifolds.  For more
details see, for example, \cite{Yano}.

An almost-K\"ahler manifold $({\mathcal M},J,\omega )$ is a real
manifold $\mathcal M$ of even dimension together with an almost-complex
structure $J$ and a symplectic form $\omega $ which are compatible
in the following sense:
\begin{equation} \omega (JX,JY)=\omega (X,Y) \,,
\end{equation}
for any vector fields $X,Y$.  In other words, the smooth field of
automorphisms
\begin{equation} J:T\mathcal M\rightarrow
T\mathcal M,\qquad J^2=-1 \label{can}
\end{equation}
is a canonical transformation of the tangent bundle w.r.t.~the
symplectic structure $\omega$, and $G(X,Y)=\omega (JX,Y)$ is
$J$-invariant (pseudo-)Riemannian metric on $\mathcal M$.

The almost-complex structure $J$ splits the complexified tangent
bundle $T^{\mathbb C}\mathcal M$ into two transverse mutually
conjugated subbundles:  $T^{\mathbb C}\mathcal
M=T^{(1,0)}\mathcal M \oplus T^{(0,1)} \mathcal M$, such that
\begin{equation}
\begin{array}{l}\displaystyle
J_pX=iX,\qquad \forall X\in T_p^{(1,0)}\mathcal M\,, \\[3mm] \displaystyle
J_pY=-iY,\quad\;\; \forall Y\in T_p^{(0,1)}\mathcal M\,,
\end{array}
\label{y}
\end{equation}
for every $p\in\mathcal M$.  In the natural frame
$\{\partial_a=\partial/\partial x^a\}$ and co-frame $\{dx^a\}$
associated to local coordinates $\{x^a\}$ on $\mathcal M$ we have
$J^a{}_b=J(dx^a,\partial_b)$, $\omega_{ab}=\omega
(\partial_a,\partial_b)$, $G_{ab}=G(\partial_a,\partial_b),$ and
\begin{equation}
J^a{}_b=G^{ac}\omega_{cb}=G_{bc}\omega^{ca},\qquad a,b,c=1,\dots,
{\rm dim}\,\mathcal M\,,
\end{equation}
where $(G^{ab})$ and $(\omega^{ab})$ are the inverse matrices to
$(G_{ab})$ and $(\omega_{ab})$, respectively.  It is known that
any symplectic manifold $(\mathcal M,\omega)$ admits a compatible
almost-complex structure $J$ turning $\mathcal M$ into an almost-K\"ahler
manifold.

Alternatively, an almost-K\"ahler manifold can be defined as a
pair $(\mathcal M,\Lambda )$ in which $\mathcal M$ is a real manifold
of even dimension
equipped with a degenerate Hermitian form $\Lambda$, such that
\begin{equation}\label{rcon}
{\rm rank\;}\Lambda =\frac 12\dim \mathcal M \,,\;\;\;\;\;\;\;\; \det
({\rm Im}\Lambda )\neq 0.
\end{equation}
The equivalence of both definitions is set by the formula
\begin{equation}
\Lambda_{ab}=G_{ab}+i\omega_{ab}. \label{lambda}
\end{equation}
Clearly, the subbundles $T^{(0,1)}\mathcal M$ and $T^{(1,0)}\mathcal M $
of the complexified tangent bundle $T^{\mathbb{C}}\mathcal M$ are
nothing but the right/left kernel distributions of the form $\Lambda$.
Note also that the left and right null-vectors of $\Lambda$ can be
obtained from each other by the complex conjugation of their components.

An almost-K\"ahler manifold $(\mathcal M,J,\omega)$ is said to
be a K\"ahler manifold if $T^{(0,1)}\mathcal M$ and
$T^{(1,0)}\mathcal M$ are integrable distributions. In this case
there exists an atlas of charts with complex coordinates
$\{z^a\}$ and holomorphic transition functions in which terms
the Hermitian form $\Lambda$ takes the block form
\begin{equation}\label{}
    \Lambda = \left(
\begin{array}{ccc}
  G_{m\bar n} & | & 0 \\
  -- & | & -- \\
  0 & | & 0 \\
\end{array}
\right)\,,\qquad m,\bar n =1,\dots,\frac12{\rm dim}\,\mathcal M\,.
\end{equation}

{}From the view point of symplectic geometry, the integrable
holomorphic/anti-holomorphic distributions $T^{(1,0)}\mathcal M$
and $T^{(0,1)}\mathcal M$ define a pair of transverse Lagrangian
polarizations of $\mathcal M$, i.e. $\omega |_{T^{(1,0)}\mathcal
M}=\omega |_{T^{(0,1)} \mathcal M}=0$. The existence of such
polarizations is of primary importance for the physical
applications as it makes possible to construct the Hilbert space
of states of a quantum-mechanical system associated to the phase
space $(\mathcal M,\omega)$.

\section{The formal K\"ahler metric construction for cotangent bundles}
\setcounter{equation}{0}

Let $({\mathcal Q}, g)$ be a (pseudo-)Riemannian manifold and let $\nabla$
be a compatible symmetric connection. For a coordinate chart
$(U,\{x^i\})\,,i=1\,\dots, \dim {\mathcal Q}$, denote by $\{p_i\}$ the
linear coordinates on the fibers of $T^\ast U$ with respect to the natural
frame $\{dx^i\}$. There is a natural lift of  the Riemannian metric $g$ on
$\mathcal Q$ to that on the total space of the cotangent bundle
$T^\ast{\mathcal Q}$. It is given by
\begin{equation}
G^{(0)}=g_{ij} dx^i\otimes d x^j+g^{ij}Dp_i\otimes Dp_j\,,
\label{1}
\end{equation}
where
\begin{displaymath}\displaystyle
g^{ik}g_{kj}=\delta^i{}_j\,,\qquad
D=dp_i\frac{\partial}{\partial p_i}-dx^i
\nabla_i\,,\qquad\nabla_i=\frac{\partial}{\partial
x^i}+p_k\Gamma_{ij}^k\frac{\partial}{\partial p_j}
\end{displaymath}
and $\Gamma_{ij}^k$ are Cristoffel symbols. Clearly,
$(T^\ast{\mathcal Q},G^{(0)})$ is a Riemannian manifold. Also,
$T^\ast{\mathcal Q}$ is a symplectic manifold w.r.t.~the
canonical symplectic structure
\begin{equation}
\omega=d\theta=Dp_i\wedge d x^i= dp_i\wedge d x^i
\,,\qquad\theta=p_i dx^i\,. \label{2}
\end{equation}
Both these structures on $T^\ast{\mathcal Q}$ can be arranged into
the Hermitian form
\begin{equation}
\Lambda^{(0)}=G^{(0)}+i\omega\,,\qquad
(\Lambda^{(0)})^\dag=\Lambda^{(0)}\,. \label{3}
\end{equation}

\vspace{5mm}\noindent {\bf Proposition 1\/}. {\it The Hermitian
form $\Lambda^{(0)}$ endows $T^\ast{\mathcal Q}$ with the
structure of an almost-K\"ahler manifold. The corresponding
almost-complex structure $J$ is integrable iff the Riemannian
manifold $({\mathcal Q},g)$ is flat.}

\vspace{5mm}\noindent {\bf Proof.\/} One can check that in each
coordinate chart $T^\ast U$ the local vector fields
\begin{equation}
{\rm Vect}(T^\ast U)\ni
{V_i}^{(0)}=\nabla_i-ig_{ij}\frac{\partial}{\partial p_j}\label{4}
\end{equation}
span the right kernel distribution of $\Lambda^{(0)}$, and this
implies the validity of the algebraic conditions~(\ref{rcon}).
Taking commutator
\begin{equation}
[{V_i}^{(0)},{V_j}^{(0)}]=R^m{}_{kij}
p_m \frac{\partial}{\partial
p_k}\,, \label{5}
\end{equation}
we see that the distribution generated by $V^{(0)}_i$ is
integrable iff the curvature tensor
\begin{equation}
R^m{}_{kij}=\frac{\partial\Gamma^m_{jk}}{\partial x^i}
-\frac{\partial\Gamma^m_{ik}}{\partial
x^j}+\Gamma^m_{in}\Gamma^n_{jk}- \Gamma^m_{jn}\Gamma^n_{ik}\,,
\label{6}
\end{equation}
of the metric $g_{ij}$ vanishes. The left kernel distribution for
$\Lambda^{(0)}$ is obtained by the complex conjugation of
$\{V^{(0)}_i\}$ and therefore it is integrable whenever
the right one is integrable. $\square$

We see that the simple ansatz (\ref{1}) for the metric on $T^\ast{
\mathcal Q}$ induces a K\"ahler structure whenever the base
Riemannian manifold ${\mathcal Q}$ is flat. If the curvature is
nonzero we can try to modify the ``bare'' metric $G^{(0)}$ by
adding higher powers in $p$'s in order to restore the
integrability of the kernel distributions.  More precisely, we
will allow the metric tensor on $T^\ast{\mathcal Q}$ to be given
by formal power series in $p$'s with smooth coefficients. The
general expression for such a metric~$G$ looks like
\begin{equation}
G=G^{(0)}+G^{(1)}+G^{(2)}+G^{(3)}+\dots\,, \label{7}
\end{equation}
where the components of the tensors $G^{(n)}$ are monomials in
$p$'s of degree $n$. Starting with $G^{(0)}$, the metric $G$ will
formally nondegenerate. We require the corresponding Hermitian
form
\begin{equation}
\Lambda=G+i\omega\,. \label{8}
\end{equation}
to have an involutory kernel distribution $\{V_i\}$ of rank
$\dim{\mathcal Q}$. As the components of the metric tensor $G$, the
components of null-vectors $V_i$ are supposed to be given by
formal powers series in $p$'s, i.e.
\begin{equation}
V_i=V^{(0)}_i-i\sum\limits_{n=1}^\infty X_{ki}{}^{j_1\dots
j_n}(x)p_{j_1}\dots p_{j_n}\frac{\partial}{\partial p_k}
=V^{(0)}_i-i\sum\limits_{n=1}^\infty X_{ki}^{(n)}\,, \label{9}
\end{equation}
so that
\begin{equation}
\Lambda(\;\cdot\;,V_i)=0\,, \label{10}
\end{equation}
and
\begin{equation}
[V_i,V_j]=0\,.\label{11}
\end{equation}

In fact, once the local vector fields $V$'s satisfying (\ref{10})
and (\ref{11}) are known, the metric $G$ is straightforwardly
restored by them.

\vspace{5mm}\noindent {\bf Proposition 2}. {\it
Given a commuting set of local vector fields $\{V_i\}$ of the
form~{\rm(\ref{9})} so that the coefficients $X_{ki}{}^{j_1\dots j_n}$
are tensors on $\mathcal Q$ being symmetric in $ki$ and $j_1\dots j_n$.
Then the $V_i$ span the right kernel distribution of the
Hermitian form $\Lambda = G+i\omega$, where
\begin{equation}
\begin{array}{c}
G=\tilde g_{ij}{}d x^i\otimes{}d x^j+\tilde g^{ij}\tilde
Dp_i\otimes \tilde Dp_j\,,
\end{array}
\label{13}
\end{equation}
and
\begin{equation}
\tilde g_{ij}=g_{ij}+{\rm Re}\,\sum\limits_{n=1}^\infty
X_{ij}^{(n)}\,,\;\;\;\;\; \tilde Dp_i={}d
p_i-\tilde\Gamma_{ji}^kp_k{}d x^j\,,\;\;\;\;
\tilde\Gamma^k_{ij}p_k=\Gamma^k_{ij}p_k+{\rm
Im}\,\sum\limits_{n=1}^\infty X_{ij}^{(n)}\,, \label{14}
\end{equation}
$\tilde g^{ij}$ being the formal inverse to the matrix $\tilde
g_{ij}$. }

\vspace{5mm}\noindent {\bf Proof.\/} Using notation (\ref{14}) one
can write
\begin{equation}
V_i=\tilde\nabla_i-i\tilde g_{ij}\frac{\partial}{\partial
p_j}\,,\qquad \tilde\nabla_i=\frac{\partial}{\partial
x^i}+\tilde\Gamma^k_{ij}p_k \frac{\partial}{\partial p_j}\,.
\label{15}
\end{equation}
Then expressions (\ref{13}), (\ref{15}) for $G$ and $V_i$ formally
coincide with those for $G^{(0)}$ and $V^{(0)}_i$. $\square$


To show that the Eq.~(\ref{11}) does have a solution satisfying
the aforementioned conditions, we start with some preparatory
constructions.

Consider the supercommutative algebra $\mathcal{ F}$ of formal
power series in $p$'s with coefficients in the exterior algebra of
differential forms on $\mathcal Q$. The general element of the algebra
$\mathcal{F}$ reads
\begin{equation} f(x,p,{}d
x)=\sum\limits_{r,q=0}^\infty f_{k_1\dots k_r}{}^{j_1\dots j_q}(x)
p_{j_1}\dots p_{j_q} d x^{k_1}\wedge\dots\wedge d x^{k_r}\,,
\label{16}
\end{equation}
where the coefficients $f_{k_1\dots k_r}{}^{j_1\dots j_q}(x)$ make
a tensor of type $(q,r)$, symmetric in the contravariant indices
and skew-symmetric in the covariant ones. Denote by
${\mathcal{F}}_{q,\,r}\subset {\mathcal{F}}$ the subspaces of
homogenous elements.

Similarly, consider the Lie superalgebra $\mathcal{D}$ of
first-order differential operators acting in $\mathcal{F}$ and
having the form
\begin{equation}
\begin{array}{c}\displaystyle
Y(x,p,d x)=\sum\limits_{r,q=0}^\infty\left( Y^m{}_{k_1\dots
k_r}{}^{j_1\dots j_q}(x) p_{j_1}\dots p_{j_q} d
x^{k_1}\wedge\dots\wedge d x^{k_r}\nabla_m+\right.\\[3mm]
\qquad\qquad\qquad\qquad\left.+\displaystyle\tilde Y_{mk_1\dots
k_r}{}^{j_1\dots j_q}(x) p_{j_1}\dots p_{j_q} d
x^{k_1}\wedge\dots\wedge d x^{k_r}\frac{\partial}{\partial p_m}\right) \,,
\end{array}\displaystyle
\label{17}
\end{equation}
where the coefficients $Y^m{}_{k_1\dots k_r}{}^{j_1\dots j_q}$ and
$\tilde Y_{mk_1\dots k_r}{}^{j_1\dots j_q}$ make tensors of type
$(q+1,r)$ and $(q,r+1)$, respectively, which are symmetric in
$j_1\dots j_q$ and skew-symmetric in  $k_1\dots k_r$. To each term
of the series (\ref{17}) we prescribe the bi-degree $(q,r)$ and
denote by ${\mathcal{D}}_{q,\,r}\subset {\mathcal{D}}$ the
subspace of all such elements. The supercommutator of two
homogeneous element from $\mathcal{D}$ is given by
\begin{displaymath}\displaystyle
[Y_1,Y_2]f=Y_1(Y_2f)-(-1)^{r(Y_1)r(Y_2)}Y_2(Y_1f)\,,\qquad\forall
f\in{\mathcal{F}}\,.
\end{displaymath}
Introduce the following two elements of the superalgebra $\mathcal{D}$:
\begin{equation}\label{18}
\begin{array}{ll} \displaystyle
\displaystyle \nabla=d x^i\nabla_i\,,&\quad\displaystyle
\delta=dx^ig_{ij}\frac{\partial}{\partial p_j}\,,
\\ \displaystyle
\nabla: \mathcal{F}_{q,\,r}\rightarrow\mathcal{F}_{q,\,r+1}\,,
&\displaystyle
\quad \delta:\mathcal{F}_{q,\,r}\rightarrow \mathcal{F}_{q-1,\,r+1}\,.
\end{array}
\end{equation}
One can check that
\begin{equation}\label{rel}
\nabla^2=\frac 12[\nabla,\nabla]=R\,,
\qquad [\nabla,\delta]=0\,,\qquad \delta^2=\frac12[\delta,\delta]=0\,,
\end{equation}
where
\begin{equation}
R=\frac12R^m{}_{kij}p_m d x^i\wedge d x^j
\frac{\partial}{\partial p_k}\,.
\end{equation}

In what follows we will need the information about the cohomology
of the nilpotent operator~$\delta$. This may be obtained by
introducing a partial homotopy operator $\delta^{-1}:
\mathcal{F}_{q,\,r}\rightarrow {\mathcal{F}}_{q+1,\,r-1}$ acting
by the rule:
\begin{equation}
\delta^{-1}f=g^{ij}p_i\left(\frac{\partial}{\partial
x^j}\right)\pint \int\limits_0^1 f(x,tp,t{}d x)\frac{{}d t}{t}\,,
\label{18a}
\end{equation}
where $Y\pint f$ stands for the contraction of the vector field
$Y$ with the form $f$. Like $\delta$, the operator $\delta^{-1}$
is nilpotent, $(\delta^{-1})^2=0$, and it satisfies the identity
\begin{equation}
\delta\delta^{-1}f+\delta^{-1}\delta
f+\pi_0f=f\,,\;\;\;\;\;\;\forall f\in \mathcal{F} \,, \label{19}
\end{equation}
where $\pi_0:{\mathcal{F}}\to{\mathcal{F}}_{0,0}= C^{\infty}({\mathcal Q})$
is the canonical projection, i.e. $\pi_0 f(x,p,d x)=f(x,0,0)$. The
last identity is similar to the Hodge-de Rham decomposition for
the exterior algebra of differential forms and says that the space
of $\delta$-cohomology coincides with $\mathcal{F}_{0,0}$.

Note that the analogous Hodge-de Rham decomposition (\ref{19})
takes place in the Lie superalgebra $\mathcal{D}$ if we put
\begin{displaymath}
\delta Y= [\delta, Y]\,, \qquad \pi_0
Y(x,p,dp)=Y(x,0,0)\,,\qquad \forall Y\in \mathcal{D}\,,
\end{displaymath}
and
define $\delta^{-1}$ by the formula (\ref{18a}) in which
$f(x,p,dp)$ is substituted by $Y(x,p,dp)$.

Consider differentiations $V\in{\mathcal{D}}_{\bullet,1}$ of the
form
\begin{equation}\label{20}
V=d x^i V_i=\nabla-i\delta-iX\,,
\end{equation}
where
\begin{equation}
X=d x^i\sum\limits_{n=1}^\infty X_{ki}{}^{j_1\dots
j_n}(x)p_{j_1}\dots p_{j_n}\frac{\partial}{\partial p_k}
=\sum\limits_{n=1}^\infty X^{(n)}\,.
\end{equation}
Clearly, the involution condition (\ref{11}) is equivalent to
\begin{equation}
V^2=\frac12 [V,V]=0\,. \label{21}
\end{equation}
In view of the obvious identities $\delta\theta=\nabla\theta=0$,
the symmetry condition $X_{ki}{}^{j_1\dots
j_n}=X_{ik}{}^{j_1\dots j_n}\,,n=1,2,\dots$, can be expressed as
\begin{equation}
V\theta=0\,,\qquad \theta=p_id x^i\,. \label{22}
\end{equation}
Thus, to construct the K\"ahler metric $G$ of the form (\ref{7}),
it suffices to find $V_i$ of the form (\ref{20}) obeying
Eqs.~(\ref{21}) and (\ref{22}).

\vspace{5mm}\noindent{\bf Proposition 3.} \textit{The general
solution to Eq.~{\rm(\ref{21})} is given by the following recurrent
formulae:
\begin{equation}
\begin{array}{l}\displaystyle
X^{(1)}=\delta a^{(2)}\,,
\\[5mm] \displaystyle X^{(2)}=\delta^{-1}
R-\delta^{-1}\left(i\nabla X^{(1)}+\frac12
[X^{(1)},X^{(1)}]\right)+\delta a^{(3)}\,,\\[5mm] \displaystyle
X^{(n)}=-\delta^{-1}\left(i\nabla
X^{(n-1)}+\frac12\sum_{p=1}^{n-1}
[X^{(p)},X^{(n-p)}]\right)+\delta a^{(n+1)}\,,\qquad n>2\,,
\end{array}
\label{25}
\end{equation}
where
\begin{equation}
a=\delta^{-1}X=\sum\limits_{n=2}^\infty\frac1n a_k{}^{j_1\dots
j_n}(x) p_{j_1}\dots p_{j_n}\frac{\partial}{\partial p_k}=
\sum\limits_{n=2}^\infty a^{(n)}
 \label{24}
\end{equation}
is an element from $\mathcal{D}_{\bullet,\,0}$.}

\vspace{5mm}\noindent
 {\bf Proof.} Using Rels.~(\ref{rel}) one finds
\begin{equation}
\frac12(V)^2= R-i\nabla X-\delta X-\frac12X^2\,, \label{23}
\end{equation}
where we have put $\nabla X\equiv [\nabla, X]$. Eq.~(\ref{21}) can
be written now as
\begin{equation}
\delta X=R-i\nabla X-\frac12X^2\,. \label{26}
\end{equation}
Applying $\delta^{-1}$ to both sides of the last equation and
accounting the Hodge-de Rham decomposition (\ref{19}) for the Lie
superalgebra $\mathcal{D}$, we get
\begin{equation}
X=\delta^{-1}R-\delta^{-1}\left(i\nabla X+\frac12X^2\right)+\delta
a\,. \label{27}
\end{equation}
Since the operator $\delta^{-1}$ increases the degree of $p$'s,
while $\nabla$ remains it intact, Eq.~(\ref{27}) can be solved by
iterations and Rels. (\ref{25}) represent the corresponding
recurrent formulae.

It remains to show that any solution $\tilde{X}$ to Eq.~(\ref{27})
obeys the initial Eq.~(\ref{26}) and the condition (\ref{24}).
First of all, the relation $\delta^{-1} \tilde{X}=a$
immediately follows from  (\ref{27}) as consequence of the
nilpotency  of the operator $\delta^{-1}$ and the Hodge-de Rham
decomposition for~$a$. Denote $\tilde{V}=\nabla-i\delta-i\tilde{X}$.
Then
\begin{equation}
\frac 12\tilde{V}^2=R-i\nabla \tilde{X}-\delta \tilde{X}-
\frac12 \tilde{X}^2\,
\end{equation}
and $\delta^{-1}(\tilde{V}^2)=0$. On the other hand, writing the Jacobi
identity  $( \tilde{V})^3=\frac14[\tilde{V},[\tilde{V},\tilde{V}]]=0$ in
the form
\begin{displaymath} \delta({\tilde{V}}^2)=(-i
\tilde{V}+\delta)\tilde{V}^2= (-i\nabla-\tilde{X})\tilde{V}^2\,
\end{displaymath}
and using the Hodge-de Rham decomposition together with
$\delta^{-1}(\tilde{V}^2)=0$, we get
\begin{equation}
\tilde{V}^2=-\delta^{-1} \left((i\nabla+
\tilde{X})\tilde{V}^2\right)\,.
\label{v22}
\end{equation}
Since the operator $(i\nabla + \tilde{X})$ does not decrease the
degree of $p$'s, while $\delta^{-1}$ increases  the degree  by
one, we conclude that Eq.~(\ref{v22}) has the unique solution
$\tilde{V}^2=0$. $\square$


To formulate the next proposition introduce the Liouville vector
field on $T^\ast{\mathcal Q}$:
\begin{equation}
{\mathcal{D}}_{1,\,0}\ni \hat N=p_i\frac{\partial}{\partial
p_i}\,. \label{28}
\end{equation}

\vspace{5mm}\noindent {\bf Proposition 4.} {\it In order for $V_i$
defined by Eq.~{\rm(\ref{25})} to satisfy the condition~{\rm(\ref{22})}
it is necessary and sufficient that the element
\begin{equation}
{\mathcal{F}}_{\bullet,\,1}\ni\tilde a=(\hat Na)\theta =
\sum\limits_{n=2}^\infty d x^k a_k{}^{j_1\dots j_n}(x)
p_{j_1}\dots p_{j_n} =\sum\limits_{n=2}^\infty \tilde a^{(n)}
\label{29}
\end{equation}
determined by Eq.~{\rm(\ref{24})} obeys the equation
\begin{equation}
V\tilde a=0\,. \label{30}
\end{equation}
The general solution to the last equation is given by the
following recurrent relations:
\begin{equation}
\begin{array}{c}\displaystyle
\tilde a^{(2)}=\delta b^{(3)}\,,\\[5mm]
\displaystyle \tilde a^{(n)}=-\delta^{-1}\left(i\nabla\tilde
a^{(n-1)}+ \sum\limits_{p=1}^{n-2} X^{(p)}\tilde
a^{(n-p)}\right)+\delta b^{(n+1)}\,,\;\;\; n\geq 3\,,
\end{array}
\label{25a}
\end{equation}
where
\begin{equation}
b=\delta^{-1}\tilde a=\sum\limits_{n=3}^\infty\frac1n b^{j_1\dots
j_n}(x) p_{j_1}\dots p_{j_n} =\sum\limits_{n=3}^\infty b^{(n)}
\label{24a}
\end{equation}
is an arbitrary element from ${\mathcal{F}}_{\bullet,\,0}$ whose
expansion starts with third order in $p$'s}.

\vspace{5mm}\noindent
 {\bf Proof.} We should analyze the condition
\begin{equation}
X\theta=0\quad\Leftrightarrow\quad X^{(n)}\theta =0\,.
\end{equation}
Clearly, $X^{(1)}\theta=0$ implies $\delta\tilde a^{(2)}=0$, and
this is equivalent to the validity of Eq.~(\ref{30}) in first
order in $p$'s. The equivalence between  Eqs.~(\ref{30}) and
(\ref{22}) can be now established by simple induction in $n$.

Making use of the formulae (\ref{25}) and the Bianchi identity
$(\delta^{-1} R)\theta=0$, one can rewrite
\begin{equation}
X^{(n)}\theta=-\delta^{-1}\left(i\nabla
X^{(n-1)}+\frac12\sum_{p=1}^{n-1}
[X^{(p)},X^{(n-p)}]\right)\theta+\delta(a^{(n+1)}\theta)=0\,,\qquad
n>1\,, \label{35}
\end{equation}
or more explicitly
\begin{equation}
\begin{array}{l}\displaystyle
X^{(n)}\theta=-\frac{1}{n+1}{}d x^m\wedge d x^k
\left[i\nabla^{j_1} X_{km}{}^{j_2\dots j_n}-i\nabla_m
X_k{}^{j_1j_2\dots
j_n}+\right.\\[3mm] \quad\quad\displaystyle\left.
+\sum\limits_{p=1}^{n-1}(n-p)
(X_s{}^{j_1j_2\dots j_{p+1}}X_{km}{}^{sj_{p+2}\dots j_n}-
X_{sm}{}^{j_1\dots j_p}X_k{}^{j_{p+1}sj_{p+2}\dots j_n})
\right]p_{j_1}\dots p_{j_n}+\delta(a^{(n+1)}\theta)\,.
\end{array}
\label{36}
\end{equation}
Here all indices are risen and lowered with the help of $g^{ij}$
and $g_{ij}$.

Proceeding by induction, we suppose that all the equations
$X^{(l)}\theta=0$ with $l<n$ are already satisfied, i.e.
$X_{km}^{(l)}=X_{mk}^{(l)}$. Then  Eq.~(\ref{36}) takes the form
\begin{equation}
X^{(n)}\theta=\frac{1}{n+1} d x^m\wedge d x^k \left[i\nabla_m
X_k{}^{j_1j_2\dots j_n}+\sum\limits_{p=1}^{n-1}(n-p)
X_{sm}{}^{j_1\dots j_p}X_k{}^{j_{p+1}sj_{p+2}\dots j_n}
\right]p_{j_1}\dots p_{j_n}+\delta(a^{(n+1)}\theta)\,.
\label{37}
\end{equation}
Notice that  Eq.~(\ref{24}) implies that
\begin{equation}
X_k{}^{j_1\dots j_n}p_{j_1}\dots p_{j_n}=a_k{}^{j_1\dots j_n}
p_{j_1}\dots p_{j_n}\,.
\label{38}
\end{equation}
Taking in account symmetry of $X$ and $a$ in contravariant
indices one can bring Eq.~(\ref{38}) into the form
\begin{equation}
(n-p)X_k{}^{j_{p+1}sj_{p+2}\dots j_n}+
X_k{}^{sj_{p+1}j_{p+2}\dots j_n}=(n-p+1)a_k{}^{sj_{p+1}j_{p+2}\dots j_n}\,.
\label{39}
\end{equation}
Substituting Eqs.~(\ref{38}) and (\ref{39}) into (\ref{37}) and
using the identity
\begin{displaymath}\displaystyle
d x^m\wedge d x^k \sum\limits_{p=1}^{n-1}X_{sm}{}^{j_1\dots j_p}
X_k{}^{sj_{p+1}j_{p+2}\dots j_n}p_{j_1}\dots p_{j_n}= 0\,,
\end{displaymath}
we get
\begin{displaymath}\displaystyle
X^{(n)}\theta=\frac{1}{n+1} d x^m\wedge d x^k \left[i\nabla_m
a_k{}^{j_1\dots j_n}+\sum\limits_{p=1}^{n-1}(n-p+1)
X_{sm}{}^{j_1\dots j_p}a_k{}^{sj_{p+1}j_{p+2}\dots j_n}
\right]p_{j_1}\dots p_{j_n}+\delta(a^{(n+1)}\theta)\,.
\end{displaymath}
Finally, comparing the last expression with Definition
(\ref{29}) we find
\begin{equation}
X^{(n)}\theta=\frac{i}{n+1} \left[\nabla\tilde
a^{(n)}-i\delta\tilde a^{(n+1)}- i\sum\limits_{p=1}^{n-1}
X^{(p)}\tilde a^{(n-p+1)}\right]= \frac{i}{n+1}(V\tilde
a)^{(n)}\,. \label{41}
\end{equation}

Returning to the Eq.~(\ref{35}) we see, that $X^{(n)}\theta=0$ is
equivalent to $(V\tilde a)^{(n)}=0$, provided all the equations
$X^{(l)}\theta=0$ with $l<n$ are satisfied, but this implies
(\ref{30}). $\square$


Thus we have shown, that the ambiguity in the recurrent definition
(\ref{25}) of the K\"ahler structure is completely described by
one function $b(x,p)$ of the form (\ref{24a}). Having a set of
formal functions  $X_{ij}(x,p)$, symmetric in indices $ij$, one
can immediately  reconstruct the K\"ahler metric $G$ by the
formula (\ref{13}).

It is also possible to rewrite $\Lambda$ in a form which makes
explicit the existence of the right/left kernel distributions. By
definition, the local vector fields (\ref{9}) generate the
K\"ahler polarization of the cotangent bundle, in particular,
\begin{equation}\label{42a}
\omega(V_i,V_j)=0\,.
\end{equation}
Alternatively, one can say that the one-forms
\begin{equation}
Z_i=-V_i\pint\omega= \tilde Dp_i+i\tilde g_{ik}d x^k =
dp_i-\Gamma^k_{ji}p_kd x^j+ig_{ij}d x^j+iX_{ji}(x,p){d x^j}
\label{42}
\end{equation}
generate the annihilator of the holomorphic distribution
$\{V_i\}$. By the Frobenius theorem, the annihilator is closed
under the exterior differentiation. A straightforward
computation yields
\begin{equation}
dZ_i=\left(-\Gamma^j_{ik}+i\frac{\partial X_{ik}}{\partial p_j}\right)
Z_j\wedge dx^k\,.
\label{44}
\end{equation}
In terms of the basis one-forms $(Z_i,\bar Z_j)$, the Hermitian form
$\Lambda$ can be written as
\begin{equation}
\Lambda=\tilde g^{ij}\bar Z_i\otimes Z_j\,.
\label{45}
\end{equation}
In view of (\ref{42a}) $V_i\pint Z_j=0$, and thus
$\Lambda(\cdot,V_i)=\Lambda(\bar V_i,\cdot)=0$.

In Ref.~\cite{DLSh} it was shown that the Hermitian form $\Lambda$
contains all the geometric prerequisites needed for the Wick
symbol construction (see also~\cite{KS}).

\section{Holomorphic coordinates
and convergence of the power series\\ for the metric}
\setcounter{equation}{0}

Since the vector fields $\{V_i\}$ (\ref{9}) give a basis in
the holomorphic distribution, we arrive at the following
definition of anti-holomorphic functions on $T^\ast\mathcal{Q}$.
The formal series
\begin{equation}
f=\sum\limits_{n=0}^\infty f^{j_1\dots j_n}(x)p_{j_1}\dots p_{j_n}
\equiv\sum\limits_{n=0}^\infty f^{(n)}\in
\mathcal{F}_{\bullet,\,0}
\label{a1}
\end{equation}
is said to be a formal anti-holomorphic function if
\begin{equation}
Vf=0\,. \label{a2}
\end{equation}

\vspace{5mm}\noindent {\bf Proposition 5.} {\it Let $f$ be a
formal anti-holomorphic function, then
\begin{equation}
df=\frac{\partial f}{\partial p_i}Z_i\,. \label{a7}
\end{equation}
where one forms $Z_i$ (\ref{42}) span the annihilator of the
holomorphic distribution $\{V_i\}$.}

\vspace{5mm}\noindent {\bf Proof.} This immediately follows from
the identity
\begin{displaymath}\displaystyle
df=\frac{\partial f}{\partial p_i}Z_i+Vf\,,\qquad V=dx^iV_i\,,
\end{displaymath}
$f$ being an arbitrary formal function on $T^\ast
\mathcal{Q}$. $\square$

Equation (\ref{a2}) can be solved iteratively in full analogy
with Eqs.~(\ref{21}) and (\ref{30}). At the $n$-th step the
solution reads
\begin{equation}
f^{(n)}=-i\delta^{-1}(\nabla
f^{(n-1)}-i\sum\limits_{r=1}^{n-1}X^{(r)}f^{(n-r)})\,,\qquad
n\geqslant 1\,, \label{a3}
\end{equation}
where $f^{(0)}(x)=f|_{p=0}$ is a given function in the coordinate
chart $(U,\{x^i\}$) and $\delta^{-1}$ is the partial homotopy
operator (\ref{18a}). If $f^{(0)n}(x)\,,n=1,\dots,{\rm
dim}\,{\mathcal Q}$, is a set of independent functions on $U$ the
corresponding anti-holomorphic functions $f^n(x,p)$ can be taken
as anti-holomorphic coordinates adapted to the formal K\"ahler
structure on $T^\ast\mathcal{Q}$.

So far we have dealt with the construction of formal K\"ahler
structure on $T^\ast \mathcal{Q}$ assuming all the ingredients are
given by formal power series in the canonical fiber coordinates.
Now, we are going to reproduce this construction from a somewhat
different, perhaps more geometrical, viewpoint. Namely, assuming
$(\mathcal{Q}, g)$ to be real-analytical Riemannian manifold, we
define an atlas of local holomorphic coordinates in a tubular
neighborhood $W\subset T^\ast\mathcal{Q}$ of the base
$\mathcal{Q}$. The coordinates are constructed as solutions of a
differential equation associated to a certain Hamiltonian flow on
$T^\ast \mathcal{Q}$ with initial data on $\mathcal{Q}$. The
complex structure, corresponding to these coordinates, turns out
to be compatible with the canonical symplectic form and thus it
defines an integrable K\"ahler structure. The latter is shown to
coincide with the formal K\"ahler structure  of  Sec.\ 3, that
implies the convergence of the power series (\ref{7}) in $W$.

The main idea behind this construction is very simple. Suppose
that $X$ satisfies the special boundary condition $\delta^{-1}X=0$
(see Eq.~(\ref{24})). Then the second term in the l.h.s.~of
Eq.~(\ref{a3}) vanishes and we have
\begin{equation}
f^{(n)}=-i\delta^{-1}\nabla f^{(n-1)}=
-\frac{i}{n}g^{ij}p_i\nabla_jf^{(n-1)}=
\frac{1}{n}\{H_0,f^{(n-1)}\}\,,\qquad n\geqslant 1\,,
\label{a3a}
\end{equation}
where
\begin{equation}
H_0=\frac{i}{2}g^{ij}(x)p_ip_j\,.\qquad
\label{a5}
\end{equation}
The formula (\ref{a3a}) implies that $f(x,p)=f(x,p;1)$, where
$f(x,p,t)$ is the solution of the equation
\begin{equation}
\frac{df}{dt}=\{H_0\,,\,f\} \label{a6}
\end{equation}
with the initial condition $f(x,p;0)=f^{(0)}(x)$. We see that the
series defined by the recurrence relation~(\ref{a3a}) is nothing
but the exponential map generated by the complex geodesic
Hamiltonian~$H_0$.

Now let us choose a set $\{f^{(0)n}\,,n=1,\dots,\dim
\mathcal{Q}\}$ of independent functions on $U$, say
$f^{(0)n}=x^n$, and let $f^n(x,p)=f^n(x,p;1)$ be their exponential
lift on $T^*\mathcal{Q}$ w.r.t. the Hamiltonian (\ref{a5}). The
standard theorems of ODE's ensure the real-analyticity of
functions $f^n(x,p)$ in a sufficiently small neighborhood $W \cap
T^\ast U$.

By construction, the complex functions $f^n(x,p)$ pair-wise
commute w.r.t.\ the canonical Poisson bracket,
\begin{equation}
\{f^m,f^n\}=0\,.
\label{a9}
\end{equation}
The K\"ahler metric is now defined as the matrix inverse of the
Poisson brackets of the holomorphic and anti-holomorphic
coordinates
\begin{equation}
G^{m\bar n}=\frac{1}{2}\{f^m,\bar f^n\}\equiv\frac{1}{2}\left(
\frac{\partial f^m}{\partial x^k}\frac{\partial\bar f^n}{\partial p_k}-
\frac{\partial\bar f^n}{\partial x^k}\frac{\partial f^m}{\partial p_k}
\right)\,.
\label{a10}
\end{equation}
Clearly, $(G^{n\bar m})$ is a tensor w.r.t. coordinate changes on
$W$.

This K\"ahler structure actually coincides with the restriction on
$W$ of the formal K\"ahler structure constructed in the previous
section. In order to see this, we express the Hermitian form
$\Lambda$ (see Eq.~(\ref{45})) in terms of the holomorphic
coordinates $f^n$.

\vspace{5mm}\noindent {\bf Proposition 6.} {\it With the above
definition we have

{\rm(i)} The matrix
\begin{equation}
e^{mi}=\frac{\partial f^m}{\partial p_i}\,,\qquad m,i=1,\dots,{\rm
dim}\,\mathcal{Q}\,,
\label{a11}
\end{equation}
is invertible for $W$ being sufficiently small.

{\rm(ii)} The one-forms
\begin{equation}
Z_i=e_{im}df^m \,,\qquad e^{mi}e_{ik}=\delta^m{}_k\,
\label{a12}
\end{equation}
are given by Eq.~{\rm(\ref{42})}, where $X_{ij}(x,p)$ are
real-analytical functions satisfying the conditions
\begin{equation}
X_{ij}(x,p)=X_{ji}(x,p)\,,\qquad X_{ij}(x,0)=0\,,\qquad
p_kg^{ki}X_{ij}(x,p)=0\,.
\label{a12pr}
\end{equation}

{\rm (iii)} The real-analytical vector fields $V_i$, defined by
the equation $V_i\pint\omega=-Z_i$, have the form~{\rm (\ref{15})} with
\begin{equation}
\begin{array}{l}\displaystyle
\tilde{g}_{ij}=\tilde g_{ji}={\rm Im}\left(\,e_{jm}\frac{\partial
f^m}{\partial x^i}\right)
=G^{k\bar m}e_{ki}\bar e_{mj}=g_{ij}+O(p^2)\,,\\[5mm] \displaystyle
\tilde{\Gamma}^k_{ij}=\tilde\Gamma^k_{ji}=-{\rm
Re}\,\left(e_{jm}\frac{\partial f^m}{\partial x^i}\right)
=\Gamma_{ij}^k+O(p)\,.
\end{array}
\label{a13}
\end{equation}

{\rm (iv)} The vector fields $V_i$ commute pair-wise and
generate a local basis of the holomorphic distribution, so that
\begin{equation}
V_i f^n=0\,.\label{a14}
\end{equation}

{\rm(v)} The Hermitian form $\Lambda$ is given by Eq.~{\rm(\ref{45})},
where $Z_j$ and $\tilde{g}^{ij}$ are defined by {\rm (ii)} and {\rm
(iii)}.}

\vspace{3mm} Given holomorphic coordinates $f^{n}$, this
proposition allows to reconstruct the re\-al-ana\-ly\-ti\-cal
$V_i$ and $\Lambda$ subject to the same defining conditions which
have been imposed on their formal counterparts from Sec.~3
with~\mbox{$\delta^{-1}X=0$}.

\vspace{5mm}\noindent {\bf Proof.} By the Cauchy theorem the
functions $f^n$ are re\-al-ana\-ly\-ti\-cal as solutions of the
differential equation (\ref{a6}) with a real-analytical r.h.s.
and the initial conditions $f^{(0)n}=x^n$. In lower orders in
$p_i$ we have
\begin{equation}
f^n=x^n-ig^{nj}p_j-\frac12\Gamma^n_{kl}g^{ki}g^{lj}p_ip_j+O(p^3)\,.
\label{a13a}
\end{equation}
and
\begin{equation}
e^{ni}=\frac{\partial f^n}{\partial p_i}=-ig^{ki}
(\delta^{n}{}_{k}+i\Gamma^n_{kl}g^{lj}p_j)+O(p^2)\,. \label{a14a}
\end{equation}
So, the matrix $(e^{ni})$ is non-degenerate for sufficiently small
$p_i$ and the inverse matrix reads
\begin{equation}
e_{in}=ig_{ik}
(\delta^{k}{}_{n}-i\Gamma^k_{nl}g^{lj}p_j)+O(p^2)\,. \label{a15}
\end{equation}
From the definition (\ref{a12}), we get
\begin{equation}
Z_i=e_{im}\left(\frac{\partial f^m}{\partial p_j}dp_j+
\frac{\partial f^m}{\partial x^j}dx^j\right)=
dp_i+e_{im}\frac{\partial f^m}{\partial x^j}dx^j\,.
\label{a16}
\end{equation}
Using (\ref{a13a}) and (\ref{a15}) one can write
\begin{equation}
e_{im}\frac{\partial f^m}{\partial x^j}=-\Gamma^k_{ij}p_k+ig_{ij}+
iX_{ij}(x,p)\,,\qquad X_{ij}(x,p)=O(p^2)\,. \label{a17}
\end{equation}
Furthermore, the involution conditions (\ref{a9}) can be brought
in the form
\begin{equation}
e_{im}\frac{\partial f^m}{\partial x^j}= e_{jm}\frac{\partial
f^m}{\partial x^i}\,, \label{a18}
\end{equation}
that implies the symmetry of $X_{ij},\tilde
g_{ij}$ and $\tilde\Gamma^k_{ij}$ in $ij$.

{}From the definition $V_i\pint\omega=-Z_i$ we find
\begin{equation}
V_i=\frac{\partial}{\partial x_i}-
e_{jm}\frac{\partial f^m}{\partial x^i}\frac{\partial}{\partial p_j}
\,,\label{a18a}
\end{equation}
and thus $V_if^m=e_{ik}\{f^m,f^k\}=0$. A direct computation
shows that $[V_i,V_j]=0$, so (iv) is proven.

The K\"ahler metric $G$ can be straightforwardly rewritten
in terms of $Z_i$
\begin{equation}
G\equiv G_{\bar nm}d\bar f^ndf^m=\tilde g^{ij}\bar Z_iZ_j\,,\qquad
\tilde g^{ij}=G_{\bar nm}\bar e^{ni}e^{mj}\,.
\label{a19}
\end{equation}
The inverse matrix to $(\tilde g^{ij})$ reads
\begin{equation}
\tilde g_{ij}=G^{n\bar m}e_{in}\bar e_{jm}=
\frac12\left(e_{im}\frac{\partial f^m}{\partial x^j}-
\bar e_{im}\frac{\partial\bar f^m}{\partial x^j}\right)=
{\rm Im}\,e_{im}\frac{\partial f^m}{\partial x^j}\,,
\label{a20}
\end{equation}
and this implies (v).

It only remains to prove that $X_{ij}g^{jk}p_k=0$ or, what is the
same, $\delta^{-1}X=0$. A solution to equation (\ref{a6}) is
nothing but the geodesic of the complex metric $ig_{ij}(x)$,
which pass, at $t=0$, through the point $\{x^i\}\in
\mathcal{Q}$ with the tangent vector  $\{g^{ij}p_j\}\in
T_x\mathcal{Q}$. Accounting for the homogeneity property of
solutions of the geodesic equation, $f^n(x,\lambda
p,\lambda^{-1}t)=f^n(x,p,t)$, $\forall \lambda \in
\mathbb{R}\backslash\{0\}$, one can write \begin{equation}
\frac{d f^n}{dt}=p_i\frac{\partial f^n}{\partial p_i}
=\{H_0\,,\,f^n\}\,, \label{a21}
\end{equation}
that yields
\begin{equation}
g^{ik}p_ke_{jn}\frac{\partial f^n}{\partial x^i}=
ig^{ik}p_k(g_{ij}-\Gamma^l_{ij}p_l)\,.
\label{a22}
\end{equation}
Comparing the last equation with~(\ref{a17}) we conclude
that $X_{ij}g^{jk}p_k=0$. $\square$


In the above consideration, we have used the complex geodesic
Hamiltonian (\ref{a5}) to define an integrable complex structure
on $W$. In fact, this construction works for an arbitrary
analytical Hamiltonian. In particular, one can consider
a Hamiltonian of the form
\begin{equation}
H=H_0+\Delta H\,,\qquad \Delta H=\sum\limits_{n=3}^\infty
h^{j_1\dots j_n}(x) p_{j_1}\dots p_{j_n} \label{a5gen}
\end{equation}
and define an atlas of holomorphic coordinates in $W$ using the
exponential map generated by the Hamiltonian flow. One can see
that the resulting K\"ahler structure corresponds to that from
Sec.~3 with nonvanishing boundary condition
$\delta^{-1}X\neq0$. The function $\Delta H(x,p)$ can be thought
of as a function corresponding to the ambiguity of the formal
K\"ahler structure described by the function $b(x,p)$
(\ref{24a}).

\section{The first Chern class}
\setcounter{equation}{0}

In the previous sections  $T^\ast{\mathcal Q}$ has been equipped
with a formal K\"ahler structure $\Lambda$ (\ref{45}).
Assuming $({\mathcal Q},g)$ to be real-analytical, we have also
shown, that the formal series for the lifted K\"ahler metric $G$
converges in a neighborhood of the zero section in
$T^\ast\mathcal Q$. In the context of deformation quantization,
it is important to identify the cohomology class of the Ricci
form $\varrho$. It is shown in Ref.~\cite{DLSh}, that the Wick-
and Weyl-type quantizations on a finite-dimensional K\"ahler
manifold are equivalent to each other iff the corresponding
Ricci form is exact, i.e.\ the corresponding cohomology class
$[\varrho]$ equals to zero. For arbitrary K\"ahler manifold
$[\varrho]$ is known to depend only on the complex structure and
it is proportional to the first Chern class of the manifold.

\vspace{5mm}\noindent {\bf Proposition 7\/}. {\it
The Ricci form of the K\"ahler structure $\Lambda$ is exact.}

\vspace{5mm}\noindent {\bf Proof.\/} Because  $T^\ast{\mathcal Q}$
(as well as any tubular neighborhood of zero section) is clearly
homotopic to the base, the canonical embedding $i:{\mathcal Q}\to
T^\ast{\mathcal Q}$ induces the isomorphism of de Rham cohomology
$H(T^\ast{\mathcal Q})\to H({\mathcal Q})$. In particular,
the Ricci form $\varrho$ of a K\"ahler structure on $T^\ast Q$ is
exact iff the same is the pull-back form $i^\ast\varrho$ on
$\mathcal Q$.

Recall the definition of the Ricci form. Let $\{x^a\}$ be a set
of local coordinates on a K\"ahler manifold, then
\begin{equation}
\varrho=\omega^{ab}K_{abcd}dx^c\wedge dx^d\,,
\label{ric1}
\end{equation}
where $\omega^{ab}$ and $K_{abcd}$ are the components of the Poisson and
Riemannian tensors associated to the K\"ahler two-form and metric
respectively.

In Sec.~3 the metric $G$ is defined by the formal
series~(\ref{14}) in the fiber coordinates, so, the Riemannian
tensor $K$ and Ricci form $\varrho$ are formal series as
well. By definition, $i^\ast\varrho$ is determined by that term
of the series $\varrho$, which does not depend on $p_i$ and
$dp_i$. A direct calculation of this term gives
\begin{equation} i^\ast\varrho=d\psi\,, \qquad
\psi=2g_{il}g_{jk} {\rm Re}\,(b^{ljk})dx^i\,, \label{ric2}
\end{equation}
where $b^{ljk}$ is the lower order term of the series~(\ref{24a}).
Hence, one can see that $i^\ast\varrho$ is exact, and therefore
\mbox{$[\rho]=0$.~$\square$}


The first Chern class of the K\"ahler structure on
$T^\ast{\mathcal Q}$ thus turns out to be zero. As mentioned
above, this implies the equivalence between the Wick and Weyl
deformation quantizations for systems with finite number
degrees of freedom. Generally speaking, this reason can not be
extended to the field theory and we turn to the discussion on this
problem in Sec.~7.

\section{Examples}
\setcounter{equation}{0}
In this section we illustrate the general construction by several examples
with special emphasis on a possible application to quantum field
theory.

{\bf 1. Cotangent bundle of a constant curvature space.} In this
case,  all the covariants of the metric $g$ are expressed
algebraically via the metric itself. In particular, the covariant
curvature tensor is given by
\begin{equation}
R_{mkij}=K(g_{mi}g_{kj}-g_{mj}g_{ki})\,,\qquad K={\rm const}\,
\label{ex1}
\end{equation}
(positive curvature corresponds to $K>0$). The metrics of
constant curvature are known also as those admitting the maximal
number of isometries. The action of the isometry group on
$\mathcal Q$ is naturally lifted to the transformation group of
the cotangent bundle $T^\ast{\mathcal Q}$. The orbits of the
latter group are given by the level sets $h={\rm const}$ of the
geodesic Hamiltonian $h=g^{ij}(x)p_ip_j$ which thus generates
the whole ring of invariant functions on $T^\ast{\mathcal Q}$.

Consider the special class of K\"ahler metrics on
$T^\ast{\mathcal Q}$ which are invariant under the action of
the isometry group. In this case the formal function $b(x,p)$,
entering to the general solution for the K\"ahler metric, should
be a function of the geodesic Hamiltonian alone, i.e.
$b(x,p)=f(h)$, $f$~being an analytical function  vanishing at
zero. Then the local vector fields $V_i$, being constructed by
the formulae (\ref{25}), will clearly have the form
\begin{equation}
V_i=\nabla_i-i(A(h)g_{im}+B(h)p_ip_m)\frac{\partial}{\partial
p_m}\,, \label{ex2}
\end{equation}
where $A(h)$, $B(h)$ are complex-valued functions of $h$,
$A(0)=1$. Taking commutators, we find
\begin{equation}
[V_i,V_j]=(2A^\prime(A+hB)-AB-K)(g_{im}p_j-g_{jm}p_i)
\frac{\partial}{\partial p_m}\,, \label{ex3}
\end{equation}
where the prime denotes the derivative in $h$. Thus, the local
vector fields $V_i$ are in involution iff
\begin{equation}
2A^\prime(A+hB)-AB-K=0\,. \label{ex4}
\end{equation}
For a fixed $B(h)$, this gives us a first-order differential
equation on $A(h)$ with initial condition $A(0)=1$. In fact,
the function $B(h)$ is uniquely determined by  $b(x,p)=f(h)$. The
recurrent procedure (\ref{25}) can  then been understood as
finding a solution for the differential equation (\ref{ex4}) in
terms of power series in the momenta~$p_i$. Given the functions
$A(h)$ and $B(h)$ satisfying Eq.~(\ref{ex4}), the K\"ahler
metric $G$ on $T^\ast{\mathcal Q}$ is defined by  Eq.~(\ref{13})
with \begin{equation} \tilde g_{ij}={\rm
Re}\,(A(h)g_{ij}+B(h)p_ip_j)\,,\quad \tilde Dp_i=d
p_i-\Gamma_{ji}^kp_k d x^j-{\rm Im}\,(A(h)g_{ij}+B(h)p_ip_j)d
x^j\,. \label{ex5} \end{equation} Choosing, for simplicity, the
functions  $A(h)$ and $B(h)$ to be real-valued, we find
\begin{equation}
G=(A(h)g_{ij}+B(h)p_ip_j)d x^i\otimes d x^j+
\frac{1}{A(h)}\left(g^{ij}-\frac{B(h)g^{im}g^{jn}p_mp_n}{A(h)+hB(h)}\right)
Dp_i\otimes Dp_j\,. \label{ex6}
\end{equation}
The general theorems for the ordinary differential equations
ensure the non-degeneracy and smoothness of $G$ in a sufficiently
small tubular neighborhood of ${\mathcal Q}$ in $T^\ast{\mathcal
Q}$ (if one identifies $\mathcal Q$ with the zero section of the
cotangent bundle). As is seen, the singularities of the metric
(\ref{ex6}) may appear as the points at which either $A(h)=0$ or
$A(h)+hB(h)=0$. From the viewpoint of the recurrent procedure, the
presence of singular points implies a finite radius of convergence
for the power series in $p$'s. To illustrate this situation,
consider three special cases when the Eq.~(\ref{ex4}) is
explicitly integrable.

i) $B=0$, $A=\sqrt{1+Kh}$,
\begin{equation}
G=\sqrt{1+Kh}\,g_{ij} d x^i\otimes d x^j+ \frac{g^{ij}Dp_i\otimes
Dp_j}{\sqrt{1+Kh}}\,. \label{ex7}
\end{equation}
When $K>0$, the K\"ahler metric  is well-defined on the whole
$T^\ast{\mathcal Q}$, whereas for  $K<0$, the singularities of the metric
form the surface $h =-K^{-1}$ . Clearly, in both cases the
radius of convergence of the power series in~$p$'s equals
$|K|^{-1}$.

ii) $B=-K$, $A=1$,
\begin{equation}
G=(g_{ij}-Kp_ip_j) d x^i\otimes d x^j+
\left(g^{ij}+\frac{Kg^{im}g^{jn}p_mp_n}{1-Kh}\right) Dp_i\otimes
Dp_j\,. \label{ex8}
\end{equation}
The situation is opposite to the previous one: the K\"ahler metric
is smooth over the whole manifold $T^\ast{\mathcal Q}$ when $K<0$, while for
$K>0$ the singular surface appears as the level set of the
geodesics Hamiltonian $h=K^{-1}$.

iii) One more interesting case when the Eq.~(\ref{ex4}) is
integrable in elementary functions corresponds to the choice
$A+hB=1$. In the language of the recurrent procedure this is
equivalent to the condition $\delta^{-1}X=0$. When $K<0$ we find
$$
A=\sqrt{|K|h}\ctg\sqrt{|K|h}
$$
and
\begin{equation}
\begin{array}{rl}
\displaystyle
G=&\displaystyle
\frac{\sqrt{|K|h}}{\tg\sqrt{|K|h}}
\left(
g_{ij}+\frac{1}{h}\left(\frac{\tg\sqrt{|K|h}}{\sqrt{|K|h}}-1\right)p_ip_j
\right)d x^i\otimes d x^j+\\ &\displaystyle+
\frac{\tg\sqrt{|K|h}}{\sqrt{|K|h}}
\left(
g^{ij}+\frac{1}{h}\left(\frac{\sqrt{|K|h}}{\tg\sqrt{|K|h}}-1\right)
g^{im}g^{jn}p_mp_n
\right)
Dp_i\otimes Dp_j\,.
\end{array}
\label{ex9}
\end{equation}
The singularities of the metric form the surface
$h=\pi^2(4|K|)^{-1}$. For $K>0$, the solution can be obtained
from~(\ref{ex9}) by simple replacement $\tg\sqrt{|K|h}\rightarrow
\th\sqrt{Kh}$. Clearly, the resulting metric appears to be
well-defined on the whole cotangent bundle $T^\ast{\mathcal Q}$.

{\bf 2. Harmonic oscillator and free scalar field.} This is the
fundamental physical model in context of which the notion of Wick
quantization has first appeared. In this case, the K\"ahler
structure is usually attributed to the representation of
creation-annihilation operators resulting from the canonical
quantization of the oscillatory variables
\begin{equation}
\begin{array}{c}
\displaystyle
a_\alpha=\frac{1}{\sqrt{2\omega_\alpha}}(P_\alpha-i\omega_\alpha
Q_\alpha)\,,\qquad\displaystyle
\bar a_\alpha=\frac{1}{\sqrt{2\omega_\alpha}}(P_\alpha+i\omega_\alpha
Q_\alpha)\,,\\[5mm]
\{a_\alpha,\bar a_\beta\}=-i\delta_{\alpha\beta}\,,\qquad
\{a_\alpha,a_\beta\}=\{\bar a_\alpha,\bar a_\beta\}=0\,,
\qquad\alpha,\beta=1,\dots,n\,.
\end{array}
\label{ex10}
\end{equation}
Here $Q_\alpha, P_\alpha$ are normal coordinates and momenta
associated to the normal frequencies $\omega_\alpha$ of an
$n$-dimensional harmonic oscillator.

{}From the geometrical viewpoint,  $a_\alpha$, $\bar a_\alpha$ are
nothing but the holomorphic and an\-ti-ho\-lo\-mor\-phic
coordinates adapted to the flat K\"ahler metric
\begin{equation}
G=\sum\limits_{\alpha=1}^n(d\bar a_\alpha\otimes da_\alpha
+da_\alpha\otimes d\bar a_\alpha)=
\sum\limits_{\alpha=1}^n
\left(\frac{1}{\omega_\alpha}dP_\alpha\otimes
dP_\alpha+\omega_\alpha dQ_\alpha\otimes dQ_\alpha \right)\,,
\label{ex11}
\end{equation}
It is instructive to rewrite this metric in arbitrary linear
canonical coordinates $x^i,p_i$. Denote by  $M=(M_{ij})$,
$K=(K_{ij})$ the corresponding matrices of mass and  stiffness,
then the Hamiltonian of the harmonic oscillator reads
\begin{equation}
H=\sum\limits_{\alpha=1}^n\omega_\alpha \bar a_\alpha a_\alpha=
\frac12(M^{ij}p_ip_j+K_{ij}x^ix^j)\,. \label{ex12}
\end{equation}
A simple linear algebra yields
\begin{equation}
G=g_{ij}dx^i\otimes dx^j+g^{ij}dp_i\otimes dp_j\,, \label{ex13}
\end{equation}
where $g=(g_{ij})$ is given by
\begin{equation}
g=M\sqrt{M^{-1}K}\,. \label{ex14}
\end{equation}
Recall that the matrix $M^{-1}K$ is diagonalizable  and its eigenvalues
coincide with the squares of normal frequencies. Thus, we see that the
K\"ahler metric on the configuration space of the harmonic
oscillator is given by expression (\ref{ex14}). In particular,
for the isotropic oscillator $K=\omega^2M$ and~$g=\omega M$.

This construction is straightforwardly  generalized to the case of
free fields, treated as a continual set of harmonic oscillators.
The free Hamiltonian of one scalar field on the (d+1)-dimensional
Minkowski space reads
\begin{equation}
\begin{array}{c}\displaystyle
H=\frac12\int d{\bf x}\,(\pi^2({\bf x})+\partial_\mu\phi({\bf x})
\partial_\mu\phi({\bf x})+m^2\phi^2({\bf x}))\,,\\ [3mm]\displaystyle
\partial_\mu=\frac{\partial}{\partial
x^\mu}\,,\quad\mu=1,\dots,d\,.
\end{array}
\label{ex15}
\end{equation}
The formal comparison with the Hamiltonian of the $n$-dimensional
oscillator suggests the following identification for the mass and
stiffness matrices:
\begin{equation}
M({\bf x},{\bf y})=\delta({\bf x}-{\bf y})\,,\qquad K({\bf x},{\bf
y})=(-\triangle_{\bf x}+m^2)\delta({\bf x}-{\bf y})\,.
\label{ex16}
\end{equation}
Correspondingly,  the infinite dimensional counterparts to the
matrix (\ref{ex14}) and its inverse are given by
\begin{equation}
\begin{array}{ll}\displaystyle
g({\bf x},{\bf y})=\sqrt{-\triangle_{\bf x}+m^2}\delta({\bf
x}-{\bf y})=&\displaystyle\int\frac{d{\bf p}}{(2\pi)^d}\,\omega
({\bf p})
e^{i{\bf p}({\bf x}-{\bf y})}\,, \\[5mm]
\displaystyle g^{-1}({\bf x},{\bf y})=\int\frac{d{\bf
p}}{(2\pi)^d}\frac{1}{\omega ({\bf p})} e^{i{\bf p}({\bf x}-{\bf
y})}\,, &\;\;\;\;\; \omega({\bf p})=\sqrt{{\bf p}^2+m^2}\,.
\end{array}
\label{ex17}
\end{equation}
The value of the K\"ahler metric  (\ref{13}) on the tangent field
$\Phi=(\delta\phi({\bf x}),\delta\pi({\bf x}))$ reads
\begin{equation}
G(\Phi,\Phi)=\int d{\bf x}d{\bf y}\, (\delta\phi({\bf x})g({\bf
x},{\bf y})\delta\phi({\bf y})+ \delta\pi({\bf x})g^{-1}({\bf
x},{\bf y})\delta\pi({\bf y}))\,. \label{ex18}
\end{equation}
Notice  that these constructions are in line with the conventional
definition of Wick symbols in the quantum field theory. Namely,
one can check that diagonal blocks  $g({\bf x},{\bf y})$ and
$g^{-1}({\bf x},{\bf y})$ of the metric (\ref{ex18}) coincide
precisely with the Wick (normal) contractions of the field
operators:
\begin{equation}
\hat\phi({\bf x})\hat\phi({\bf y})-:\hat\phi({\bf x})\hat\phi({\bf
y}): =\frac12 g^{-1}({\bf x},{\bf y})\,,\qquad \hat\pi({\bf
x})\hat\pi({\bf y})-:\hat\pi({\bf x})\hat\pi({\bf y}): =\frac12
g({\bf x},{\bf y})\,. \label{ex19}
\end{equation}
In terms of holomorphic coordinates
\begin{equation}
a({\bf p})= \int\frac{d{\bf
x}}{(2\pi){\lefteqn{{}^{\!\frac{d}{2}}}}} \,\frac{e^{-i{\bf
px}}}{\sqrt{2\omega({\bf p})}} (\pi({\bf x})-i\omega({\bf
p})\phi({\bf x}))\,,
\end{equation}
adapted to the K\"ahler metric (\ref{ex18}), the Hamiltonian
(\ref{ex15}) takes the form
\begin{equation}
H=\int d{\bf p}\,\omega({\bf p})\bar a({\bf p})a({\bf p})\,.
\end{equation}
Upon quantization the fields $\bar a({\bf p}), a({\bf p})$ turn to
the standard creation/annihilation operators.

{\bf 3. Nonlinear models.} Consider now the general Hamiltonian
describing a ``natural mechanical system''. This is given by the
sum of kinetic and potential energies:
\begin{equation}
H=\frac12h^{ij}(x)p_ip_j+V(x)\,. \label{ex20}
\end{equation}
As usual, the kinetic term defines (and is defined by) some metric
$h$ on the configuration space of the model $\mathcal Q$. Using
this metric one can immediately define the K\"ahler structure on
$T^\ast{\mathcal Q}$ by the general procedure of Sec.~3, just
identifying $h$ with $g$ in the ``bare'' metric (\ref{1}). This
option, however, seems to be not so natural or, at least, it is
not the only choice possible. A simple comparison with the case of
harmonic oscillator suggests to identify $(h_{ij})$ with the mass
matrix $M$ rather than with a genuine configuration space metric
$g$. The identification for the stiffness matrix $K$ is not so
evident and it bears a great amount of ambiguity. The most general
ansatz, compatible with the general covariance and harmonic
approximation, is given by the following expression:
\begin{equation}
K_{ij}(x)=\nabla_i\nabla_jV(x)+O_{ij}(R,V,\nabla{R},\nabla V,
\nabla^2{ R},\dots)\,. \label{ex21}
\end{equation}
Here $\nabla_i$ is a covariant derivative compatible with $h$ and
$O_{ij}$ stands for all possible non-minimal terms depending on
the curvature $R$ of the metric $h$ and vanishing in the flat
limit. The metric~$g$ is given by the same expression as for the
harmonic oscillator (\ref{ex14}) with substitution $M=h$;
in doing so, the matrix $h^{-1}K$ is supposed to be positive
definite.

The field-theoretical counterpart of the above Hamiltonian is
known as nonlinear sigma-model. The configuration space of the
model consists of all (smooth) maps $\phi:
\mathbb{R}^{d,1}\rightarrow {\mathcal Q}$ from the $(d+1$)-dimensional
Minkowski space to a Riemannian manifold $\mathcal Q$ with the metric
$h_{ij}$. In terms of linear coordinates $x=({\bf x},t)$ on
$\mathbb{R}^{d,1}$ and local coordinates $\phi^i$ on the target
manifold~$\mathcal Q$, any such map is given locally by functions
$\phi^i(x)$, which are considered as a set of scalar fields on
$\mathbb{R}^{d,1}$. The Hamiltonian of the sigma-model has the
standard form $H=T+V$, where
\begin{equation}
T[\phi,\pi]= \frac12\int d{\bf x}\, h^{ij}(\phi)\pi_i\pi_j
\,,\qquad V[\phi]=\frac12\int d{\bf x}\, h_{ij}(\phi)
\partial_\mu\phi^i\partial_\mu\phi^j\,,
\label{ex23}
\end{equation}
and $\pi_i({\bf x})$ are the momenta canonically conjugate to
$\phi^i({\bf x})$,
$\{\phi^i({\bf x}),\pi_j({\bf y})\}=\delta^i{}_j\delta({\bf x}-{\bf y})$.
The mass matrix associated with the functional of kinetic energy
is given by
\begin{equation}\label{mmm}
M_{ij}({\bf x},{\bf y})=h_{ij}(\phi({\bf x}))\delta ({\bf x}-{\bf
y})\,.
\end{equation}
As we have argued above,  the simplest choice for the generalized
stiffness matrix $K$  is given by the second covariant derivative
of the functional  $V[\phi]$.  So we put
\begin{equation}
\begin{array}{l}\displaystyle
K(\delta \phi,\delta\phi)=\int d{\bf x}d{\bf y}\,K_{ij}({\bf
x},{\bf y}) \delta\phi^i({\bf x})\delta\phi^j({\bf y})=\\[5mm]
\displaystyle  =\int d{\bf x}d{\bf y}\left[\frac{\delta^2V[\phi]}
{\delta\phi^i({\bf x})\delta\phi^j({\bf y})}- \int d{\bf
z}\Gamma^l_{ij}({\bf z}, {\bf x},{\bf y})\frac{\delta V[\phi]}
{\delta\phi^l({\bf z})} \right]\delta\phi^i({\bf
x})\delta\phi^j({\bf y})\,.
\end{array}
\label{ex24}
\end{equation}
Here
\begin{equation}
\Gamma^l_{ij}({\bf z},{\bf x},{\bf y})=\gamma^l_{ij}(\phi ({\bf
z}))\delta({\bf z}-{\bf x})\delta({\bf z}-{\bf y})
\end{equation}
and
\begin{equation}
\gamma^l_{ij} (\phi) = \frac12h^{lk}(\phi) \left(\frac{\partial
h_{kj}(\phi)}{\partial\phi^i}+ \frac{\partial
h_{ik}(\phi)}{\partial\phi^j}- \frac{\partial
h_{ij}(\phi)}{\partial\phi^k}\right)
\end{equation}
are the Cristoffel symbols of the metrics $M_{ij}({\bf x},{\bf
y})$ and $h_{ij}(\phi)$, respectively. A simple computation yields
\begin{equation}
K(\delta\phi,\delta\phi) = \int d{\bf x}\,( h_{ij}(\phi){
D}_\mu\delta\phi^i{D}_\mu\delta\phi^j-{
R}_{ikjl}(\phi)\partial_\mu\phi^k
\partial_\mu\phi^l\delta\phi^i\delta\phi^j)\,,\label{ex26}
\end{equation}
where
\begin{equation}
D_\mu\delta\phi^i=\partial_\mu\delta\phi^i+\gamma^i_{kj}
(\phi)\partial_\mu\phi^k\delta\phi^j \,\label{ex25}
\end{equation}
is the covariant derivative and ${R}_{ikjl}(\phi)$ is the
curvature tensor of $h_{ij}(\phi)$. Notice that in the flat limit
the second term in the expression (\ref{ex26}) for the generalized
stiffness matrix $K$ vanishes and, in principle,  it can be
included into the non-minimal terms in (\ref{ex21}).

\section{Conclusion}

In this paper we suggest a method of equipping any cotangent
bundle over Riemannian manifold with a formal K\"ahler structure.
Having the K\"ahler structure at hands one can perform an
explicitly covariant Wick deformation quantization of a wide class
of physical systems along the lines of the general scheme of Ref.
\cite{DLSh}. With this regard several questions may appear.

First, there is a question of equivalence between the Wick and
Weyl deformation quantizations on $T^\ast\mathcal{Q}$. The vanishing
of the first Chern class, shown in Sec.\ 5, implies that such an
equivalence does take place when $\mathcal Q$ is an ordinary manifold. This
formal equivalence, however, should be taken cautiously in the
field-theoretical context as the equivalence transform from Wick
to Weyl symbols may actually diverge because of infinite number of
degrees of freedom\footnote{Upon regularization this
transformation can provide no equivalence anymore. Well known
example of this phenomenon is offered by the Wick and Weyl
quantizations of the free bosonic string, where the phase space is
linear and no global obstructions are possible anyway. It seems
that in most physically relevant field theories Wick symbols are
not equivalent to the Weyl ones.}. So, in order to quantize a
realistic physical theories one should work with the Wick symbols
from the very beginning.

The second question concerns the fact that the K\"ahler metric is
constructed in a form of power series in momenta, so the
$\ast$-products between generic phase space functions can be
evaluated only as power expansions. This fact does not seem to be
an actual restriction for the field/string theory where most
physically interesting  classical observables are polynomial in
momenta. For example, a typical problem is computing the
$\ast$-square of the BRST charge involving first class constraints
which almost always are at most squared in momenta.

The third problem to be mentioned in relation to possible
field-theoretical applications concerns ``the right choice'' for
the metric on the configuration space of fields. In this paper we
have proposed a simple ansatz for such a metric
$g=h\sqrt{h^{-1}K}$, where $h$ is the Hesse matrix defined by
functional of kinetic energy and $K$ is the matrix of second
covariant  derivatives (w.r.t.~$h$) of the potential term. Being
quite natural and compatible with the experience of free models
(where it reproduces the standard Wick symbols in the linear phase
space), this choice is by no means unique as one may add any
expressions vanishing in the flat limit like the second term in
(\ref{ex21}) or (\ref{ex26}). One may hope, however, that under
certain physical restrictions (like invariance under the global or
local symmetries, spatial locality, etc.) all such redefinitions
of the metric will lead to formally equivalent quantum theories.
On the other hand, the formal equivalence, as has been mentioned
above, can be broken in the field theory by regularization of the
divergencies, so one can not exclude existence of inequivalent
Wick symbols generated by different choices of the configuration
space metric.

We are planing to address these questions later.

\section*{\protect\large Acknowledgements}
Authors are thankful to R.~Marnelius, D.~Tsimpis for useful
discussions at early stage of this work. This research was
supported in part by RFBR under the grants 03-02-17657 and
03-02-06709, Russian Ministry of Education under the grant E
02-3.1-250, by the INTAS grant 00-00262 and by the grant
for Support of Russian Scientific Schools 1743.2003.2.
AAS appreciates the financial support of the Dynasty Foundation and the
International Center for Fundamental Physics in Moscow.


\end{document}